# TIME/COMPUTATIONALLY OPTIMAL NETWORK ARCHITECTURE: WIRELESS SENSOR FUSION

Gadi GayathriDevi[1]
Deepika[1]
Priya Kumari[1]
Eslavath Jyoshna[1]
Dr.Garimella Rama Murthy[2]
[1]Department of Computer Science
Indian School Of Mines, Dhanbad
[2]Communications Research Lab,
International Institute of Information and Technology, Hyderabad

**ABSTRACT:**

**In this research paper, the problems dealing with sensor network architecture, sensor fusion are addressed. Time/Computationally optimal network architectures are investigated. Some novel ideas on sensor fusion are proposed**.

## I. INTRODUCTION

A **Wireless Sensor Network** consists of a large number of sensor nodes deployed in sensor field to sense the physical and environmental conditions and forward the data either in single hop or multiple hops to the base station [1]

Each sensor node consists of: a sensing unit, a processing unit, a transceiver unit and a power unit.The analog signal from the environment is digitalized using **ADC** and then sent to the controllers in the **processing unit** and **transceiver unit** provides communication between node and base station. Power is inversely proportional to the square of the distance, i.e. more amount of energy is required to send data in a single hop by a sensor node and if multiple hops are used less amount of power for each sensor is dissipated. Hence, multiple hops are of greater use than single hops if the distance is greater than one unit. It is the major issue in WSN because once the nodes are placed in their respective places replacement is costly and often difficult in inaccessible deployment regions. Therefore, the nodes are designed in such a way that whenever the sensor node is idle it immediately goes to sleep mode and activate only if it detects some information or data. However, it consumes more power while switching from sleep mode to active mode. **Power unit** manages the consumption of power [1]

*Routing algorithms:*

Flooding: It is a simple routing technique developed for multi hop networks where a node after receiving the data broadcasts the data to all its neighbours, and the process continues at other intermediate nodes until the data reaches the base station. Signal interference and congestion are the major limitations.

To overcome the drawbacks in flooding, *controlled flooding* was developed where a node after receiving the data instead of broadcasting, randomly picks the sensor nodes and transfers the data. This picking is probabilistic. [2]

## II. RELATED LITERATURE

Wireless Sensor Network is different from other wireless network in a sense that there is In-Network computation (distributed computation) i.e. locally at the sensors, cluster heads and globally at the base stations.

Data obtained from each sensor node may contain uncertainty and thus it would be more reliable to take into account the readings from various sensor nodes. In other words, combination of data derived from various sensor nodes would provide more accurate and useful result, similar to humans who combine signals from five sensory organs to make decision about their present and future actions.

**Sensor fusion**: Sensor fusion is the combining of sensory data from disparate sources such that the resulting information is in some sense better than that would be possible when these sources were used individually.

Some sensed values are crisp and some are interval valued. Therefore, there can be two approaches for fusion:
- Conversion of all values to crisp values and fuse them.
- Conversion of all values to interval valued reading and fuse them.

**Crisp Fusion functions:**

Mean/Average, Median, Mode/Maximum, Minimum are uniquely defined crisp fusion functions. Depending upon applications, different crisp fusion functions are used. For example, in case of fire detection, maximum function is used, minimum function is used to detect the temperature in areas of snowfall, etc.

**Interval Fusion Functions:**

Due to **uncertainty of readings, the idea of tolerance is taken into consideration,** thereby converting crisp readings into intervals. The tolerance considered may or may not be uniform i.e. left tolerance may or may not be equal to right tolerance. Also, the tolerance at different sensors could be different.

*Fault Tolerant fusion function:*

1. Marzullo's Algorithm(M-function): If there are 'n' intervals and atmost 'f' of them are faulty then according to this algorithm, intersection of 'n-f' intervals may contain true value. It can be implemented in O(*nlogn*) computational time by using sorting of input intervals.[3]

2. Schmid-Schossmaier Algorithm(S-function): Let a set $I=\{I_1,I_2,I_3......I_n\}$ of n number of intervals with atmost f of them are faulty. Here $I_i=[x_i,y_i]$, then output estimate is $L=[X,Y]$ where
   $X = (f+1)$th largest of the left edges $x_1,x_2,x_3 ... x_n$.
   $Y = (f+1)$th smallest of the right edges $y_1, y_2, y_3....y_n$.

   This algorithm is Lipschitz continuous and have the same computational complexity ($O(nlogn)$) as the previous algorithm[3]

3. Prasad-Iyengar'sAlgorithm( $\Omega$ -function/overlap function): Overlap function in tamely faulty sensors would create high and wide peaks in comparison of wildely faulty sensors which produce smaller and narrower peaks. This omega function results in an integration with the highest peak and the widest spread at the certain resolution.[1]

4. N-Function:The N function improves the $\Omega$ function to only generate the interval with the overlap function ranges [n-f, n]. This function is able to reduce the width of the output interval in most cases and produces a more narrower output interval by reducing the width when there are large number of sensors involved[1]

## III .Innovative Ideas on Sensor Fusion

**Binary readings:** There may be a case where the sensor node is designed in such a way that it outputs either 0(NO) or 1(YES). Here the fusion can be done using 'majority logic'. For ex, if in a building there are 5 sensor nodes arranged for detecting fire and if output is 01101 then according to the majority logic the output is assumed to be 1 and corresponding action is taken. The binary readings from sensors are fused at the cluster head stage, over the multiple cluster heads and at the base station. Thus, the fusion function can be an arbitrary Boolean function.

**Incorporating ideas of interval fusion functions into crisp fusion functions :**

Assume that there are 'n' crisp readings in which atmost 'f' are faulty. The crisp readings are sorted either in increasing or decreasing order and then remove the right f faulty readings and take the average of remaining (n-f) readings, say q. similarly, remove the left f faulty reading and take the average of remaining (n-f) readings, say p. the fused value would be the mid point of the interval either [p,q] or [q,p]

**IV .Time Optimal,Computationally Optimal Network Architecture**

A general Distributed sensor network (DSN) consists of a set of sensor nodes, processing elements (PEs), and a communication network interconnecting the various PEs. One or more sensors is associated with each PE. One sensor can report to more than one PE. A PE and its associated sensors are referred to as a cluster. Data are transferred from sensors to their associated PE(s), where the data integration takes place. PEs can also coordinate with each other to achieve a better estimation of the environment and report to higher level PEs[2]

**Main goal**: Computing Sensor Fusion functions such that
1. Computational Complexity is minimum (no. of additions, no. of comparison etc.)
2. Time complexity is minimum( minimize delay in communication from sensors to base station).
Specifically we consider fusion function concentrating mainly on crisp readings.

In some applications, the controllable aspect of architecture is sensor connections and the data structure representing the connectivity of sensors and base station.

**Lemma 1:** The computation complexity of min, max, mean fusion functions cannot be less than O(n) operations.

*Proof*: It is easy to see that no matter how we organize the nodes, the computation of minimum, maximum functions takes atleast (n-1) comparisons. Similarly, mean computation, by its very definition takes atleast(n-1) additions and one division. Therefore, computational complexity would not be less than O(n) operations.

One way to *minimize the complexity of median* can be done by using Binary Search Trees. If there are n readings then it is obvious that median would be (N/2)th value if N is even and (N+1)/2th value if N is odd. Now, readings are arranged in BST which is done in complexity of O(logn) then traverse the tree in inorder. (time complexity of inorder traversal is O(n)).

Selection of [n/2]th value gives the median.

T(n)= T(arranging in BST) + T(in-order traversal)+T( searching [n/2]th term)

T(n)= O(logn) +O(n)+O(n/2)

T(n)= O(n)< O(nlogn)

Now, it can be concluded that complexity of many symmetric fusion functions is O(n) which cannot be minimized further.

Depth of Binary tree(d) = $\log_2(n+1)-1$.
Delay = Depth of BT
Therefore, delay is of O(logn).

We are not only interested in deciding the data structure (connecting wireless sensor nodes) that minimizes computational complexity but also the time complexity.

**Lemma 2:** If each link ( edge in data structure) is associated with a delay of 'd' units then the Hub and Spoke data structure leads to the minimum possible time delay ( time complexity)

*Proof:* Among all possible data structures which represent a connected graph, the Hub-and-spoke architecture leads to a time delay of 'd' units. It is clear that no other connectivity structure on 'N' nodes can have a delay smaller than d units.

**Lemma 3:** The 'Line connectivity' data structure among 'N' nodes has maximum time complexity i.e. (N-1)d units.

*Proof:* When all the sensor nodes are arranged linearly with each link having a time delay of 'd' units then it is clear that the total time delay is (N-1)d units.

Thus, the above two lemmas fix the lower/upper bounds on time complexity.

On considering only tree data structure it is reasonably clear that with various types of tree with associated depth, the time complexity lies in between the above lower and upper bounds.

**Lemma 4:** If all the nodes are organized into a binary tree, with base station as the root node then the total time delay = $\log_2(N+2)$.

Proof:
$2^1+2^2+2^3+\ldots 2^d=N$
$1+2^1+2^2+2^3+\ldots 2^d=N+1$
$2^{d+1}-1=N+1$
$2^{d+1}=N+2$
$(d+1)\log_2 2 = \log_2(N+2)$
$d+1= \log_2(N+2)$
$\Rightarrow d= \log_2(N+2)-1$
i.e. depth of binary tree $d= \log_2(N+2)$.

**Lemma 5:** If all the nodes are organized into a q-ary tree with base station as the root node then the total time delay is $\log_q(q(N+1)-N)-1$.

Proof:
$q+q^2+q^3+\ldots q^d=N$
$1+ q+q^2+q^3+\ldots q^d=N+1$
$(q^{d+1}-1)/(q-1)=N+1$
$q^{d+1}=q(N+1)-N$
$d+1=\log_q (q(N+1)-N)$
$d= \log_q(q(N+1)-N)-1$

Thus, it is observed that on increasing the value of 'q' (i.e. the no. of children) the time complexity decreases. It is least for Hub and Spoke data structure where q=(N-1) (i.e. d=O(1)). However, it is less reliable and impractical and do not cover large area. So network implementation via binary tree data structure is more realistic and reasonable.

Relation between **sensorfusion** and **networkarchitecture**

Sensor fusion is dependent on network architecture. In fusing all sensor readings in minimum time computationally optimal algorithm is used. To send the fused value from the sensor nodes to the Base station through cluster heads in minimum possible time the time optimal algorithm is used.

**Remark:** It should be noted that the results on network architecture apply to all situations where a graph represents e.g. wireless nets on a local line.

**V .Network Architecture:Graph Entropy**

**Relation of entropy with the complexity of graphs:**

Entropy $H(x) = -\sum p_i \log_2 p_i$

In the research literature, there are many notions of graph entropy based on a probability mass function defined on the vertices. In [5], one such probability mass function over vertices is proposed.

Suppose $p_i$ where i= 1 to M is the probability mass function of discrete random variable. Then, for a graph $p_i$ can also be calculated using the expression:

$p_i$= (degree of ith node in the graph)/(total degree of the graph)

Entropy depends only upon probability mass function but not upon the values assumed by a random variables.

It is well known that Entropy becomes maximum when $p_i=1/M$ for all $1 \leq i \leq M$ and minimum when probability is 1 for exactly one outcome and probability for all other outcome is 0.

**Claim:** Graph has maximum entropy when all vertices have same degree.

**Lemma 6:** The graph entropy of ring and clique is maximum.

Proof: Let us consider a ring on n vertices. It is clear that each node has a degree 2. Hence $p_i$ for all $1 \leq i \leq n$ is same. Hence it can be stated from corollary 1 that ring has maximum entropy.

**Note:** There are other graphs on N vertices which also have maximum entropy.

**Lemma 7:** Among all possible trees on n vertices, line connected tree has maximum entropy and hub-and-spoke has minimum entropy.

Proof: Entropy of a graph would be maximum if degree of each vertex is equal (i.e. for ring and clique). Similarly, in a tree, entropy would be maximum for equal vertex degree distribution. But this type of tree is not possible, therefore, only the possible tree having maximum entropy would be line connected since each node has same degree except a root and a leaf node.

Proof by Induction : The above statement can also be proved using induction. If n=1, 2 and 3 the only possible way to connect the nodes is line connected.

For N=4, the nodes can be arranged in either line connected or tri-nary tree. Using the expression of entropy, it is clear that entropy of line connected (i.e. 1.918) is greater than the entropy of tri-nary tree(i.e. 1.7925) which states that line connected has maximum enropy for n=4.

Let us assume that it is true for n=k.

To prove that it is true for k+1 we connect the k+1 th node to the tree obtained by k nodes. These n nodes can be arranged in any number of different tree data structures in which one is line connected. When the k+1th node is connected to these trees, we get new different number of trees in which one tree will be a line connected and as it is true for n nodes it is also true for k+1 nodes that line connected has maximum entropy. Hence it is proved by induction that line connected has maximum entropy.

Entropy is minimum if the p.m.f. = {1,0,0,0…….} i.e one vertex has the maximum degree and the remaining has degree 0 each. It is possible only when the tree is disconnected. Thus, the next best case would be to give one node the maximum possible degree (i.e. N-1) and the remaining nodes with minimum possible degree (i.e. 1), which results in a hub-and-spoke.

**Lemma 8:** For a balanced binary tree on n vertices (of depth d) the graph entropy is $1/x[3(2^{d-1}-1)\log_2 3 + 2\log_2 2] - 1/x[(5*2^{d-1}-1)\log_2 x]$.

Proof: In a binary tree all intermediate nodes have degree 3 , root node have degree 2 and all leaf node have degree 1. Total degree x = $2 + 3(2^{d-1}-1) + 2^d$ . Hence $p_i$ for root node is

2/x, for intermediate nodes 3/x and for leaf nodes 1/x.

$H(x) = -\sum p_i \log_2 p_i$
$H(x) = -[2/x \log_2(2/x) + (2^{d-1}1)(3/x \log_2(3/x)) + 2^d(1/x \log_2(1/x))]$
$H(x) = 1/x[3(2^{d-1}-1)\log_2 3 + 2\log_2 2] 1/x[(5*2^{d-1}-1)\log_2 x]$. ( where $x = 2 + 3(2^{d-1}-1) + 2^d$ ).

**REMARK:** The results of graph theory can be applied in any field but here it has been used in wireless sensor networks.

## IV. CONCLUSION

In this research paper, time as well as computationally optimal data structures (dealing with network architecture) are investigated and results are derived. The data structures are evaluated using a notion of graph entropy. Some innovative ideas on sensor fusion are proposed.

## VII .REFERENCES